\documentclass[usenatbib,useAMS]{eas}
\usepackage{pslatex,graphicx}
\usepackage{amsmath,amsfonts}
\usepackage{pdfscreen}

\def\idm#1{{\mbox{\scriptsize #1}}}
\newcommand{\Hfull}{\mathcal{H}}
\newcommand{\Hsec}{\mathcal{H}_{\idm{sec}}}
\newcommand{\Hrel}{\mathcal{H}_{\idm{GR}}}

\newcommand{\Hpert}{\mathcal{H}_{\idm{pert}}}
\newcommand{\Hmut}{\mathcal{H}_{\idm{NG}}}

\newcommand{\HKepl}{\mathcal{H}_{\idm{kepl}}}

\newcommand{\AMD}{\mbox{AMD}}

\begin{document}

%%-----------------------------
%%      the top matter
%%-----------------------------
\title{Relativistic Lidov-Kozai resonance in binaries} 
\runningtitle{Relativistic Lidov-Kozai resonance}
\author{Cezary Migaszewski}
\address{Toru\'n Centre for Astronomy, Nicolaus Copernicus University,
Gagarin Str. 11, 87-100 Toru\'n, Poland;
\email{[c.migaszewski,k.gozdziewski]@astri.umk.pl}}
\author{Krzysztof Go\'zdziewski}
\sameaddress{1}
\begin{abstract}
We consider the secular dynamics of a binary and a planet in terms of
non-restricted, hierarchical three-body problem, including the general
relativity corrections to the Newtonian gravity.  We determine regions in the
parameter space where the relativistic corrections may be important for the
long-term dynamics. We try to constrain the inclinations of putative Jovian
planets in recently announced binary systems of HD~4113 and HD~156846.
\end{abstract}
\maketitle
%
%-------------------------------------------------------------------------------
\section{Introduction}
%-------------------------------------------------------------------------------
In the recent sample of detected extrasolar planetary systems, some planets
exhibit large eccentricities. It may be explained by the Lidov-Kozai resonance
(LKR) acting in binary stellar systems (e.g., \cite{Innanen1997},
\cite{Takeda2005}, \cite{Verrier2008}). If the
inner planetary orbit is inclined to the orbital plane of the binary, the
exchange of the angular momentum between orbits may force large amplitude
eccentricity oscillations of the planetary orbit, and simultaneously its
argument of pericenter $\omega_1$ librates around $\pm \pi/2$. However, the LKR
may be suppressed by the general relativity (GR) correction to the Newtonian
gravity (NG) through changing frequencies of pericenters. Here, we focus on
the non-restricted problem and relatively compact systems, and the dynamical
effects of including the GR interactions in the model of motion.
%
%-------------------------------------------------------------------------------
\section{The secular dynamics of the hierarchical triple system}
%-------------------------------------------------------------------------------
%
We consider the hierarchical triple system. The Hamiltonian  written with
respect to canonical Poincar\'e variables (e.g., \cite{Laskar1995}),
$
\Hfull = \HKepl + \Hpert,
$
where
\begin{equation}
\HKepl = \sum_{i=1}^{2} {\bigg( \frac{\mathbf{p}_i^2}{2 \beta_i} 
- \frac{\mu^*_i \beta_i}{r_i} \bigg)}, \qquad 
\Hpert = {\bigg(
 - \frac{k^2 m_1 m_2}{\Delta} +
\frac{\mathbf{p}_1 \cdot \mathbf{p}_2}{m_0}\bigg)} + {\Hrel},
\label{HKepl}
\end{equation}
describes perturbed Keplerian motions of the inner binary (the central mass
${m_0}$ and $m_1$), and the outer binary ($m_0$ and more distant point-mass
$m_2$), ${\mu^*_i=k^2~(m_0+m_i)}$, where $k$ is the Gauss gravitational constant,
${\beta_i=(1/m_i+1/m_0)^{-1}}$ are the reduced masses, ${\mathbf{r}_{1,2}}$, are
the radius vectors of  $m_{1,2}$ relative to $m_0$, $\mathbf{p}_{1,2}$ stand for
their conjugate momenta relative to the {\em barycenter}, and
${\Delta=\|\mathbf{r}_1-\mathbf{r}_2\|}$. $\Hrel$ stands for GR correction  to
the Newtonian  potential of $m_0$ and $m_1$ (see, e.g.,
\cite{Richardson1988}). We assume that the ratio of semi-major axes
$\alpha=a_1/a_2<0.2$, and $\Hpert \ll \HKepl$. It means that both $m_{1,2}$ are
small ({\em planetary regime}) or one of $m_{1,2} \sim m_0$ is relatively large,
and one of these bodies is enough distant from $m_0$ ({\em binary regime}).  

\begin{figure}
\centerline{
\vbox{
    \hbox{\includegraphics[width=4cm]{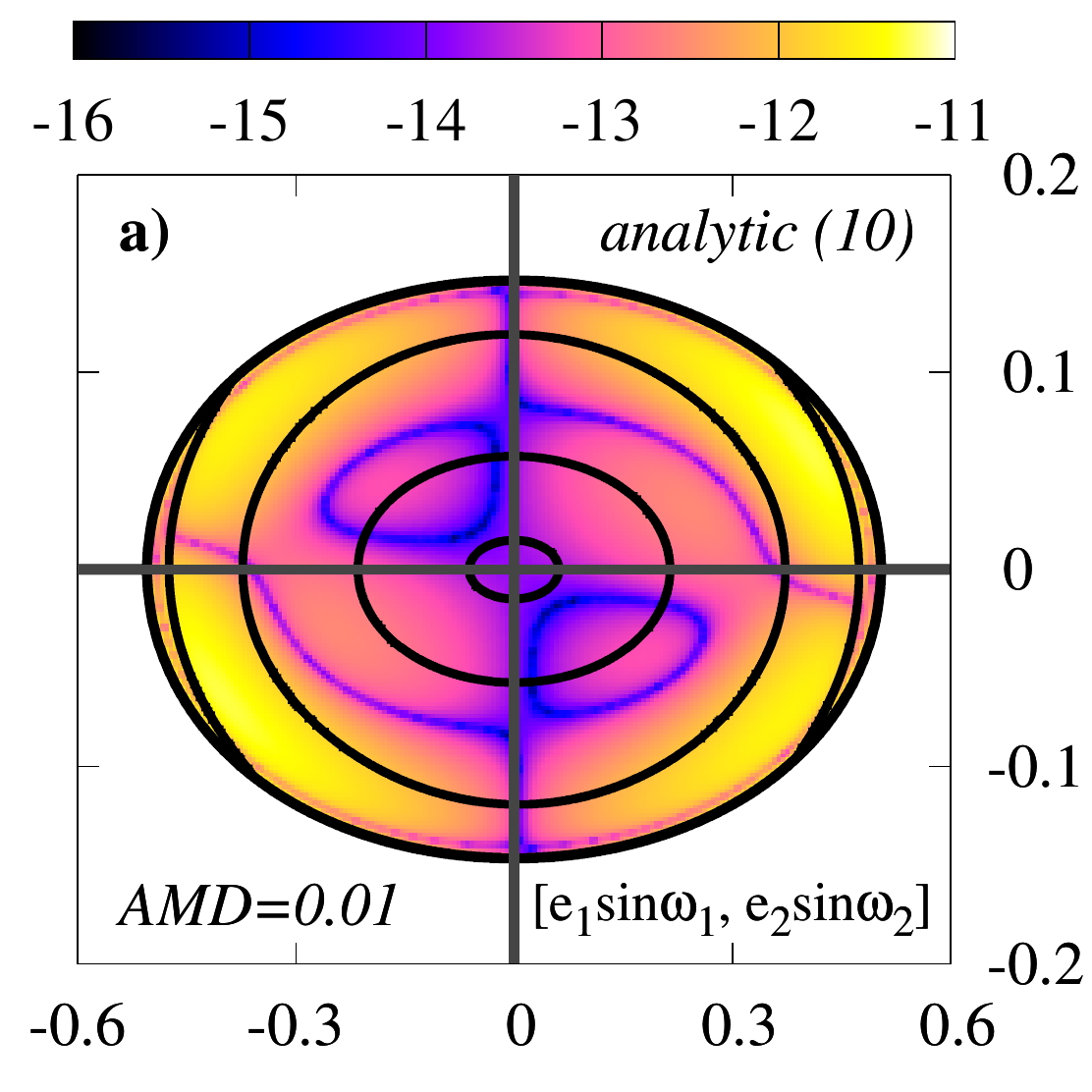}
          \includegraphics[width=4cm]{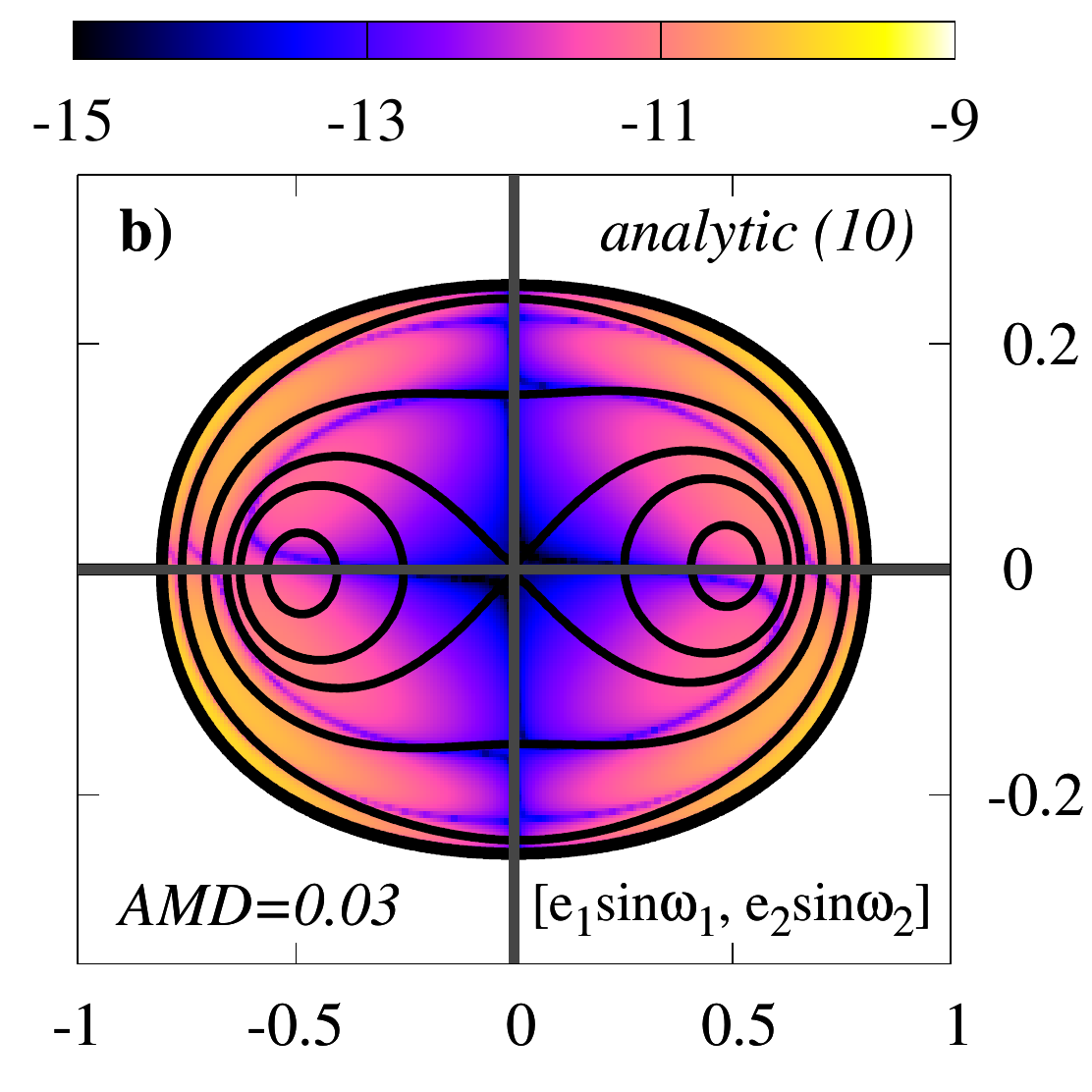}
          \includegraphics[width=4cm]{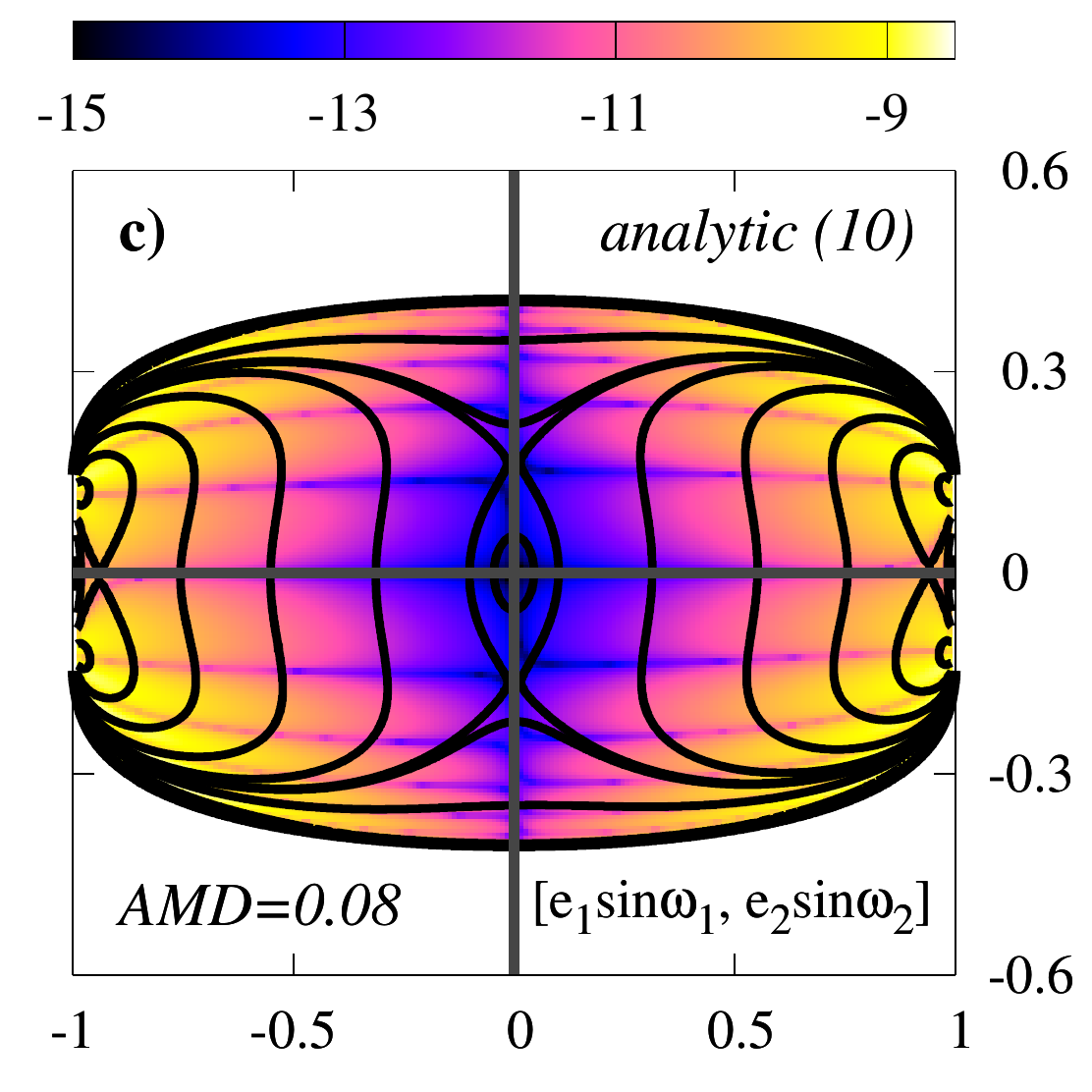}}
         }}
\caption{
A test of the relative accuracy of the 10-th order expansion of $\Hsec$, and
levels of $\Hsec$ in the $\Sigma_S$-plane (see the text for details) for
different values of $\AMD$ compared with the semi-analytical (exact) averaging
(see \cite{Michtchenko2004} or \cite{Migaszewski2008c}). Differences between the
theories are expressed in terms of the relative log-scale.
}
\label{accuracy}
\end{figure}

We expand $\Hmut$ with respect to $\alpha$ and the Hamiltonian 
is averaged out with respect to the mean
longitudes (\cite{Migaszewski2008a}), that leads to the secular term
$\Hsec = \left<\Hmut\right> +\left<\Hrel\right>$, where
\begin{equation}
\left<\Hmut\right> = -\frac{k^2 m_1 m_2}{a_2}
\left[1 + \sqrt{1-e_2^2} \sum_{l=2}^{\infty}
{\mathcal{X}^l
\mathcal{R}_l(e_1,e_2,\omega_1,\omega_2,I)}\right], 
\ {\mathcal X}=\alpha/(1-e_2^2),
\label{Hpert}
\end{equation}
$I$ stands for the mutual inclination, $\omega_{1,2}$ are the pericenter
arguments, and perturbing terms ${\mathcal R}_l$ are derived in (Migaszewski \&
Go\'zdziewski, in preparation). The averaged GR term is
$\left<\Hrel\right> = -3\beta_1\mu_1^2 c^{-2} a_1^{-2} (1 - e_1^2)^{-1/2}$,
where $c$ is the velocity of light. The expansion in Eq.~\ref{Hpert} generalizes
the octupole theory (e.g., \cite{Ford2000}) and the coplanar model
(\cite{Migaszewski2008a}). After the Jacobi's elimination of nodes
($\Delta\Omega=\pm\pi$), we eliminate one degree of freedom thanks to the
integral of the total angular momentum, $\bf C$. Then $\Hsec \equiv
\Hsec(G_1,G_2,\omega_1,\omega_2)$  parameterized by $C = |\bf C|$  (or the
Angular Momentum Deficit $\AMD \equiv L_1 + L_2 -C$, where $L_{1,2},G_{1,2}$ are
the Delaunay actions) is reduced to {\em two degrees of freedom}. For
$\alpha=0.1$,  the relative errors of $\Hsec$ approximated by the 10-th order
expansion do not exceed $10^{-8}$ in the relative magnitude (see
Fig.~\ref{accuracy}).

To study $\Hsec$, we apply the {\em representative plane of initial
conditions}, $\Sigma$, introduced in (\cite{Michtchenko2004},
\cite{Michtchenko2006}) which crosses all phase-space trajectories.  Due to
symmetries of $\Hsec$ with respect to the apsidal and nodal lines:
\begin{equation}
\frac{\partial{\Hsec}}{\partial{\omega_1}} =
\frac{\partial{\Hsec}}{\partial{\omega_2}} = 0, \ (\omega_1, \omega_2)
\in \lbrace\,(0,0),\,(0,\pm\pi),\,(\pm\pi/2,\pm\pi/2),\,
(\pm\pi/2,\mp\pi/2)\,\rbrace,
\label{symetry}
\end{equation}
and these conditions define the $\Sigma$-plane,
$
\Sigma = 
\lbrace  e_1 \cos{\Delta\varpi}, e_2 \cos{2\omega_1}\rbrace 
$, 
$\Delta\varpi \equiv \varpi_1 - \varpi_2$, 
$(e_1, e_2) \in [0,1)$, see the left-hand  panel of Fig.~\ref{Icrit_cl} for an
illustration. Restricting ($\omega_1,\omega_2$) to the above set, we also
define 
$
\Sigma_S =  \lbrace e_1 \sin{\omega_1}, e_2 \sin{\omega_2}\rbrace 
$, $
\Sigma_C =  \lbrace e_1 \cos{\omega_1}, e_2 \cos{\omega_2} \rbrace
$
revealing
levels of $\Hsec$ without discontinuities (\cite{Libert2007}).

%
%-------------------------------------------------------------------------------
\section{Stationary solutions and the Lidov-Kozai resonance}
\begin{figure}
\centerline{
    \hbox{\includegraphics[width=5.8cm]{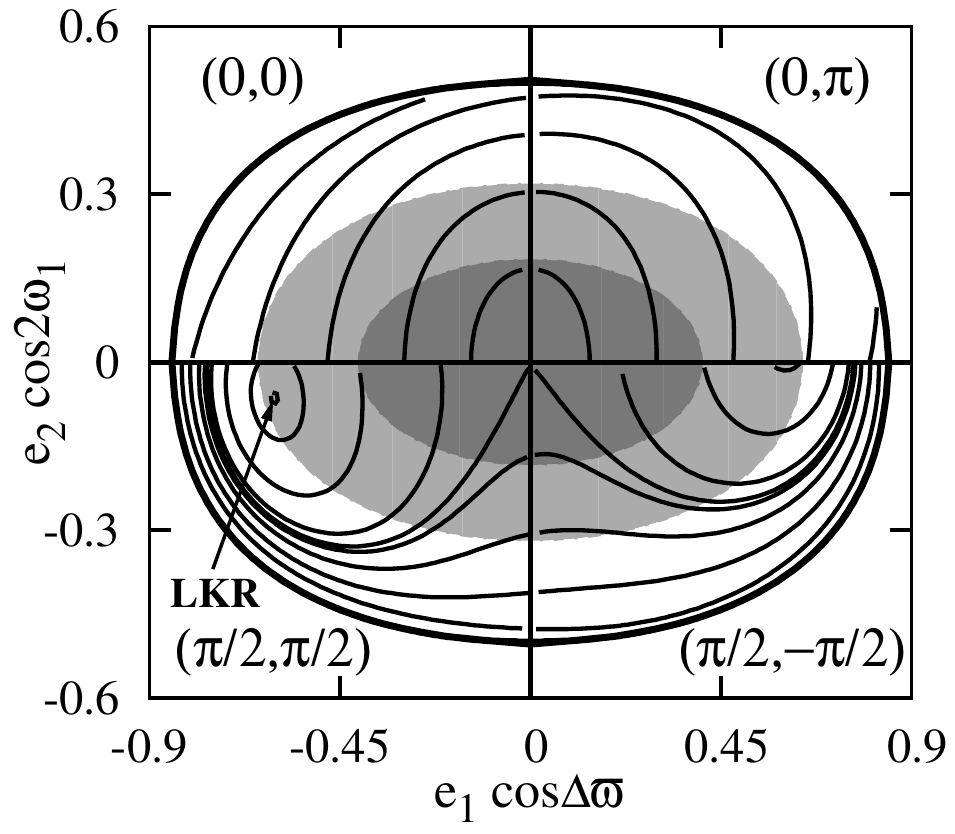}\hspace*{5mm}
          \includegraphics[width=5.1cm]{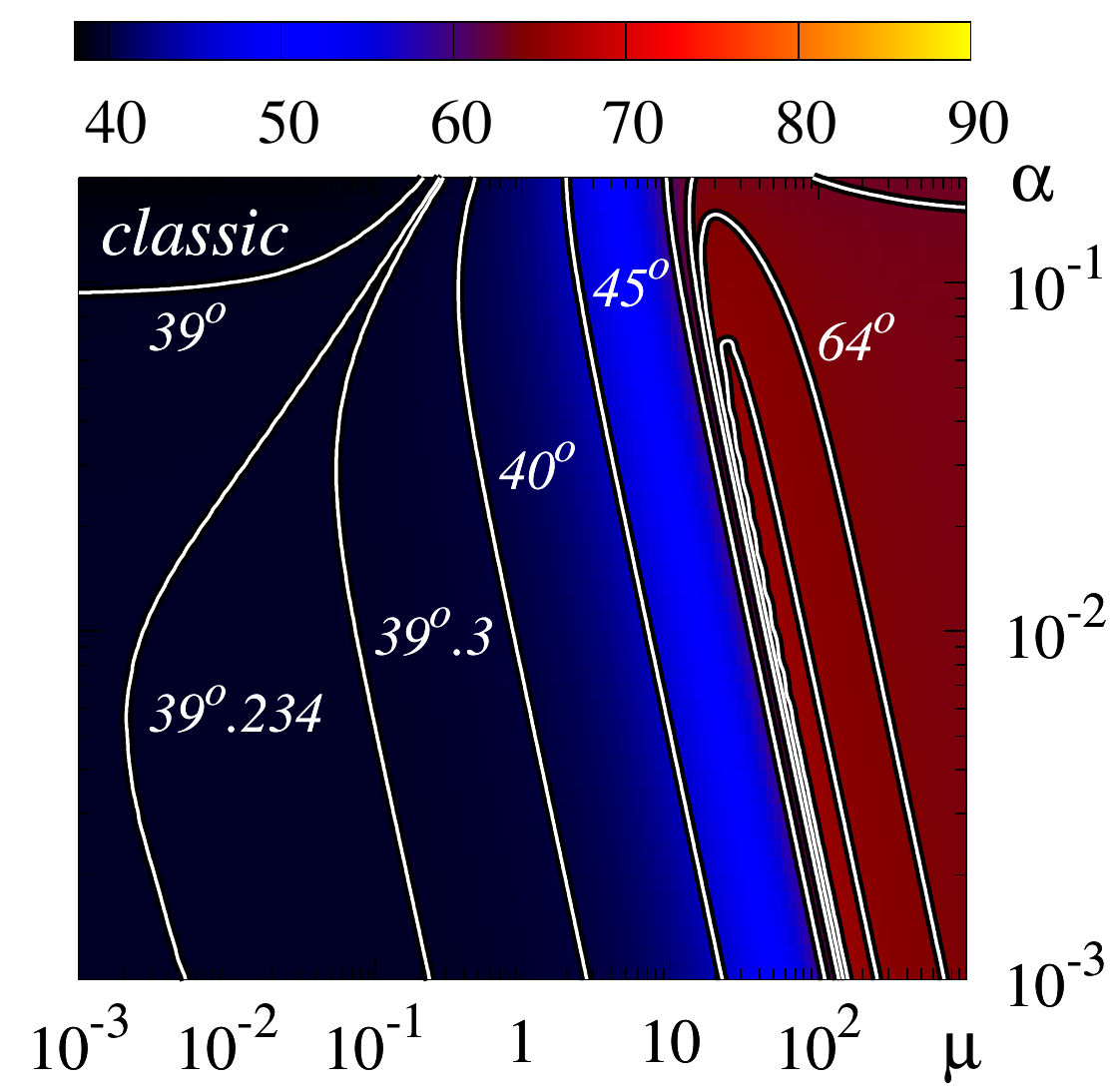}
}
}
\caption{{\em The left-hand panel:} the representative plane of initial
conditions, $\Sigma$. {\em The right-hand panel}: the critical inclination
$i_{\idm{crit}}$ in the $(\mu,\alpha)$-plane, the NG model. See the text for
more details.}
\label{Icrit_cl}
\end{figure}

%-------------------------------------------------------------------------------
%
The equilibria of $\Hsec$ provide much information on the structure of
the phase space. In the $\Sigma$-planes, these equilibria appear as
quasi-elliptic or quasi-hyperbolic (saddle) points of the levels of
$\Hsec$,
according with the equations of motion:
\begin{equation}
\frac{\partial{\Hsec}}{\partial{G_1}}=0, \quad 
\frac{\partial{\Hsec}}{\partial{G_2}}=0, \qquad \mbox{or} \qquad
\frac{\partial{\Hsec}}{\partial{e_1}} = 0, \quad
\frac{\partial{\Hsec}}{\partial{e_2}} = 0.
\label{equilibria}
\end{equation}
The stability and bifurcations of equilibria in the full and in the restricted
three-body problem were studied in many works (see, e.g., \cite{Kozai1962},
\cite{Krasinsky1972}, \cite{Krasinsky1974}, \cite{Lidov1974}, \cite{Fejoz2002}, 
\cite{Michtchenko2006}, \cite{Libert2007},  \cite{Migaszewski2008c}) regarding
the NG model. Here, we investigate more closely the equilibrium at the origin
($e_1=e _2=0$), which is well known since Poincar\'e, in the presence of the GR
interactions. According with the terminology of \cite{Krasinsky1974}, that is
the trivial {space solution of the 3rd kind} ($e=0,I\neq 0$), see
Fig.~\ref{accuracy}a. The zero-eccentricity equilibrium (ZEE) is related to {\em
the maximum} of $\Hsec$ and is Lyapunov stable. For a given value of  $C$, the
mutual inclination of circular orbits, $i_0$, is also a maximal mutual
inclination if $I_{1,2}<\pi/2$.  Moreover, for some smaller $C$ (larger $\AMD$),
the origin may change its stability due to bifurcations illustrated in the
$\Sigma_S$-plane (Figs.~\ref{accuracy}b,c). For instance,   Fig.~\ref{accuracy}b
illustrates a saddle accompanied by two elliptic points.  Close to these points,
the phase-space trajectories exhibit librations of $\omega_{1,2}$ around $\pm
\pi/2$.  This structure (see also Fig.~\ref{Icrit_cl}) is associated with the
LKR; the elliptic points may be called  {nontrivial, negative solutions of the
3rd kind}, ($e\neq 0, I\neq 0$), as in (\cite{Krasinsky1974}). They appear when
$C<C_\idm{crit}$, or, equivalently, when $i_0>i_\idm{crit}$ for initially {\em
circular} orbits. For more details see e.g., (\cite{Libert2007},
\cite{Migaszewski2008c}). 

Here, we restrict our calculations to $i_\idm{crit}<\pi/2$ (the case of {\em
direct orbits}),  hence we do not follow the second bifurcation
(Fig.~\ref{accuracy}c) appearing for $i_\idm{crit} \sim \pi/2, e_1 \sim 1$.  We
compute  $i_\idm{crit}$ causing the stability change of ZEE in the NG model for
mass ratio $\mu \equiv m_1/m_2  \in [10^{-3},10^{3}]$, and $\alpha \in
[10^{-3},0.2]$ (see the right-hand panel of Fig.~\ref{Icrit_cl}). Two kinds of
LK bifurcation  may appear (\cite{Krasinsky1972}):  at $i_\idm{crit} \sim
40^{\circ}$ ({the inner LKR}, amplifying $e_1$) and for $i_\idm{crit} \sim
64^{\circ}$  (we call it {\em the outer LKR}; e.g, in a case of a circumbinary
planet).   Moreover, in the  planetary regime of $m_{1,2}$,  $i_\idm{crit}$
{depends only on $\alpha$ and $\mu$}, and not on individual semi-major axes nor
masses.
%
%-------------------------------------------------------------------------------
\section{Effects of the General Relativity correction}
\begin{figure}
\centerline{
 \vbox{
    \hbox{\includegraphics [width=4cm]{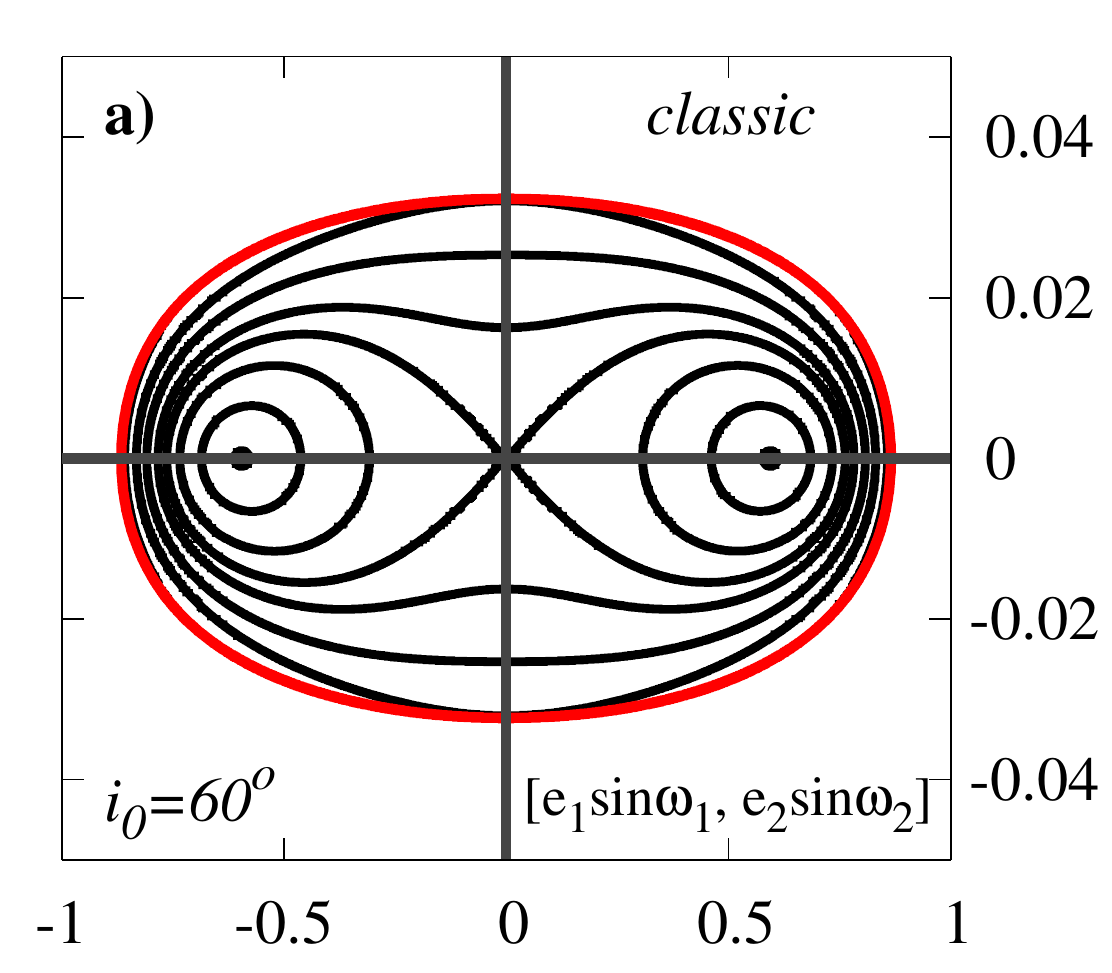}
          \includegraphics [width=4cm]{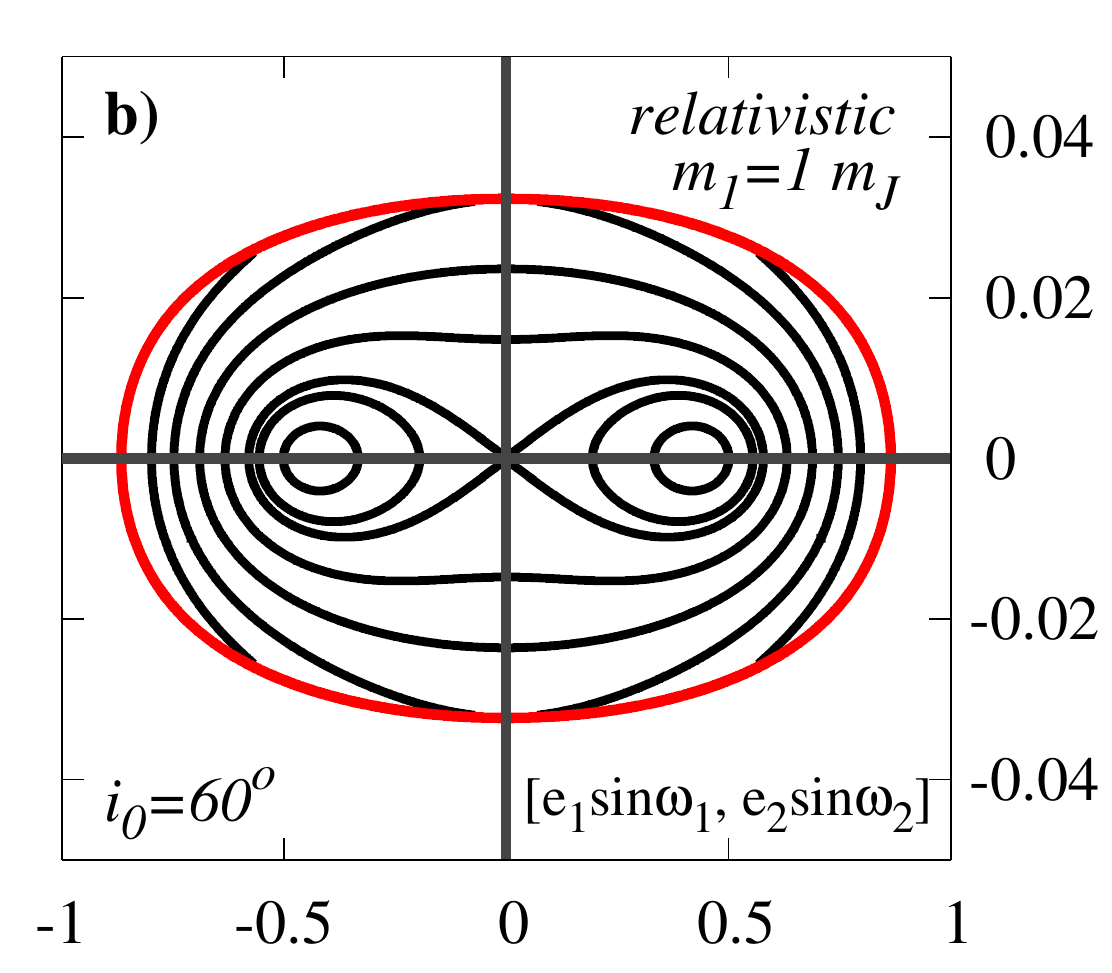}
          \includegraphics [width=4cm]{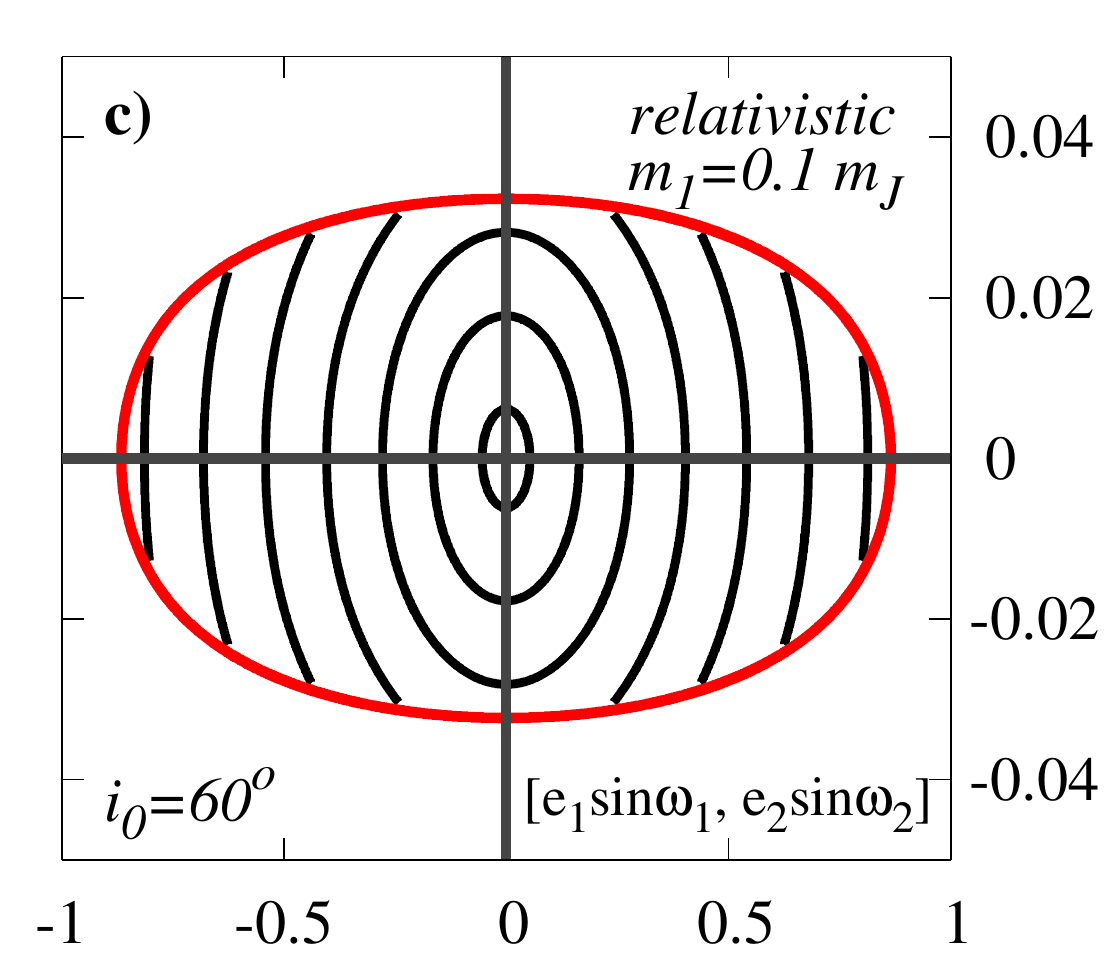}
	  }
         }}
\caption{
The $\Sigma_s$-plane for $m_0=1\mbox{m}_{\odot}$, $a_1=0.5\mbox{au}$,
$\alpha=0.01$, $\mu=0.01$, and $i_0=60^{\circ}$. The left-hand panel is for the
NG model, next panels are for the GR model, and 
$m_1=\{1,0.1\}{\mbox{m}}_\idm{J}$, respectively.
}
\label{rel_rep}
\end{figure}

\begin{figure}
\centerline{
\vbox{
    \hbox{\includegraphics [width=4cm]{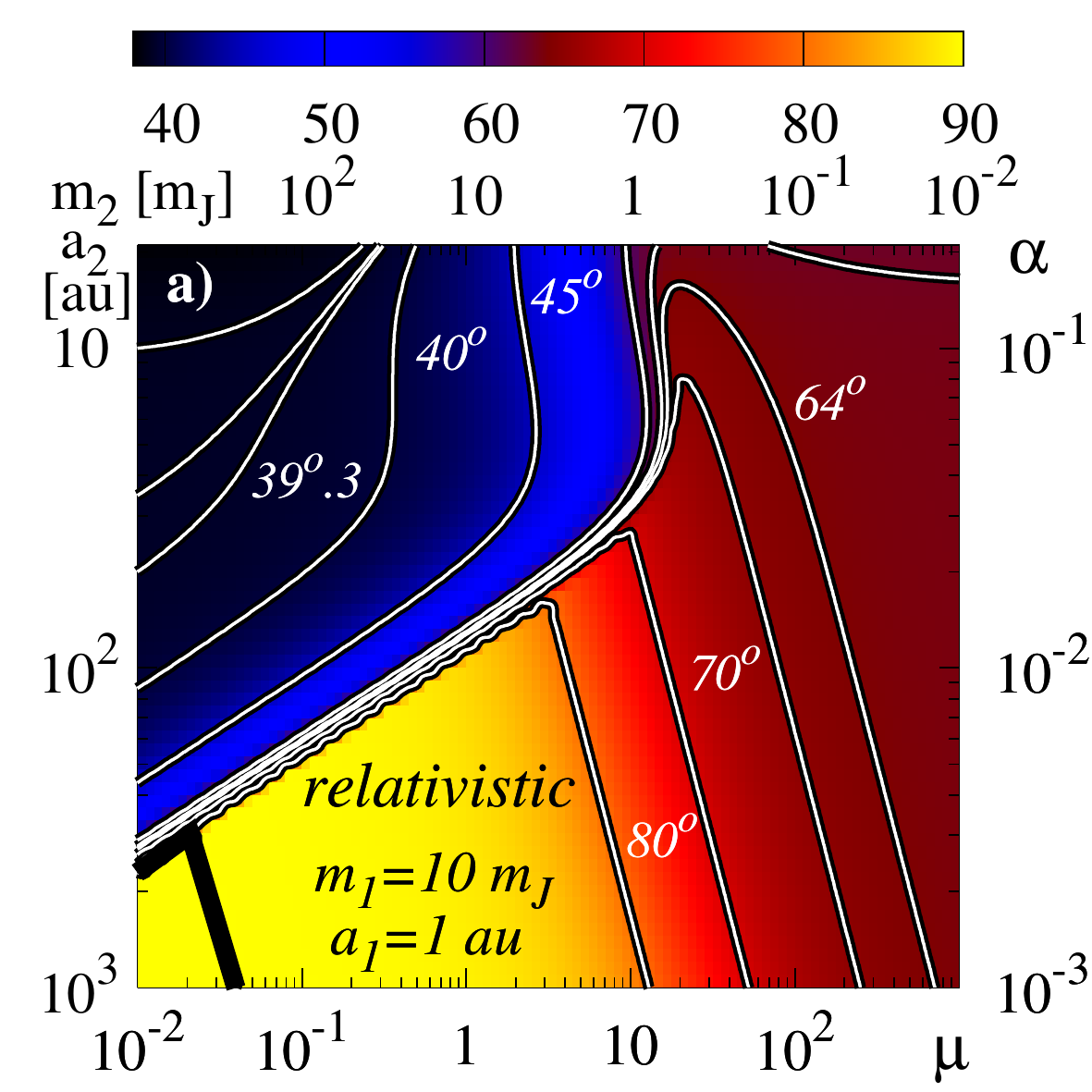}
          \includegraphics [width=4cm]{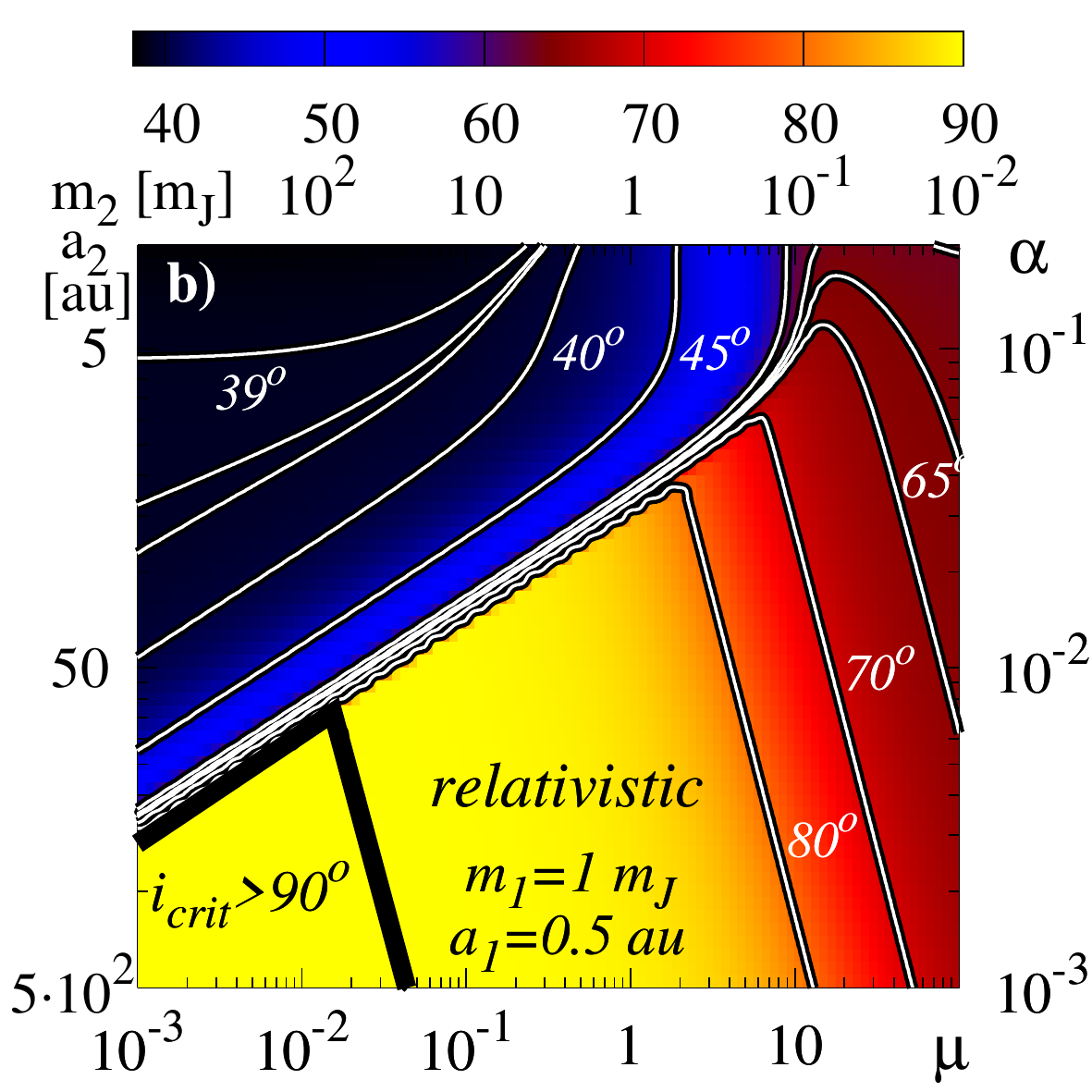}
          \includegraphics [width=4cm]{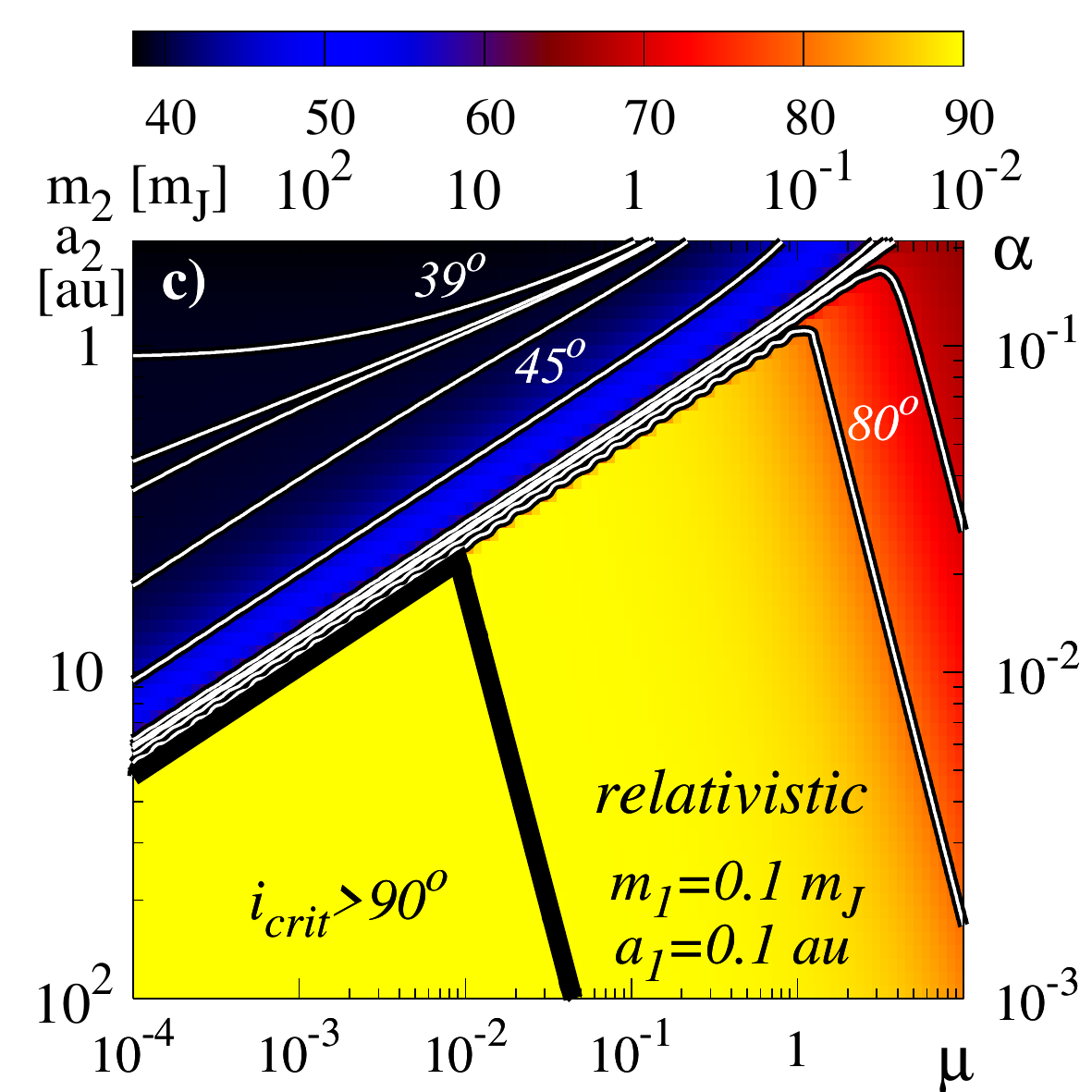}
	 }
     }
}
\caption{
The critical inclination $i_{\idm{crit}}$ in the $(\mu,\alpha)$-plane, in terms
of the relativistic model.   Orbital parameters of the innermost body are
labeled in the respective plots, $m_0=1\mbox{m}_{\odot}$.
}
\label{Icrit_rel}
\end{figure}

%-------------------------------------------------------------------------------
%
After introducing the $\Hrel$ correction to $\Hsec$, the structure of the phase
space changes qualitatively (Fig.~\ref{rel_rep}). We choose  the same $i_0$ 
(a function of constant $L_{1,2}$ and $C$) for
fixed $\alpha=0.01$ and $\mu=0.01$. For the NG model, a clear LKR structure
appears (Fig.~\ref{rel_rep}a). However, in terms of the GR model, the saddle
structure may shrink (Fig.~\ref{rel_rep}b), and finally, for small enough
masses, it disappears (Fig.~\ref{rel_rep}c). That effect may be characterized
globally  through the critical inclination $i_\idm{crit} \equiv 
i_\idm{crit}(\mu,\alpha)$ for varying $m_1, a_1$
(Fig.~\ref{Icrit_rel}). The structure  of the $(\mu,\alpha)$-plane in terms of
the GR model is very different from the NG case (Fig.~\ref{Icrit_cl}). We may
see {three} distinct  regions related to the inner LKR  ($i_\idm{crit} \sim
40^{\circ}$), and to the outer LKR ($i_\idm{crit} \sim 64^{\circ}$), smoothly
passing into {\em a new region} emerging in the bottom-left corner, which is 
colored in yellow, where $i_\idm{crit} \rightarrow \pi/2$, and the LKR may be
totally suppressed. 

%
%-------------------------------------------------------------------------------
\section{An application to the HD~4113 and HD~156846 planetary systems}
\begin{figure}
\centerline{
\vbox{
    \hbox{\includegraphics [width=4cm]{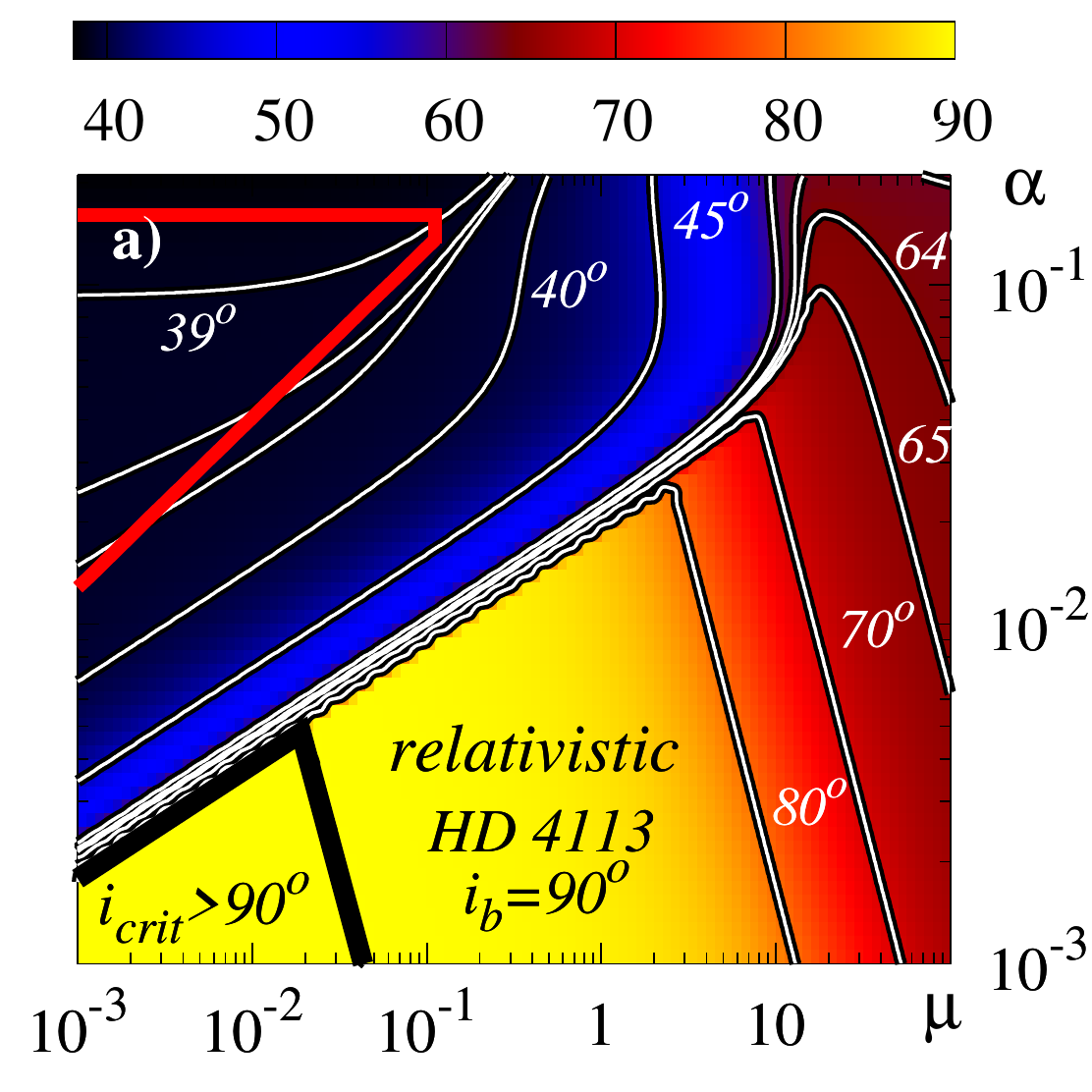}
          \includegraphics [width=4cm]{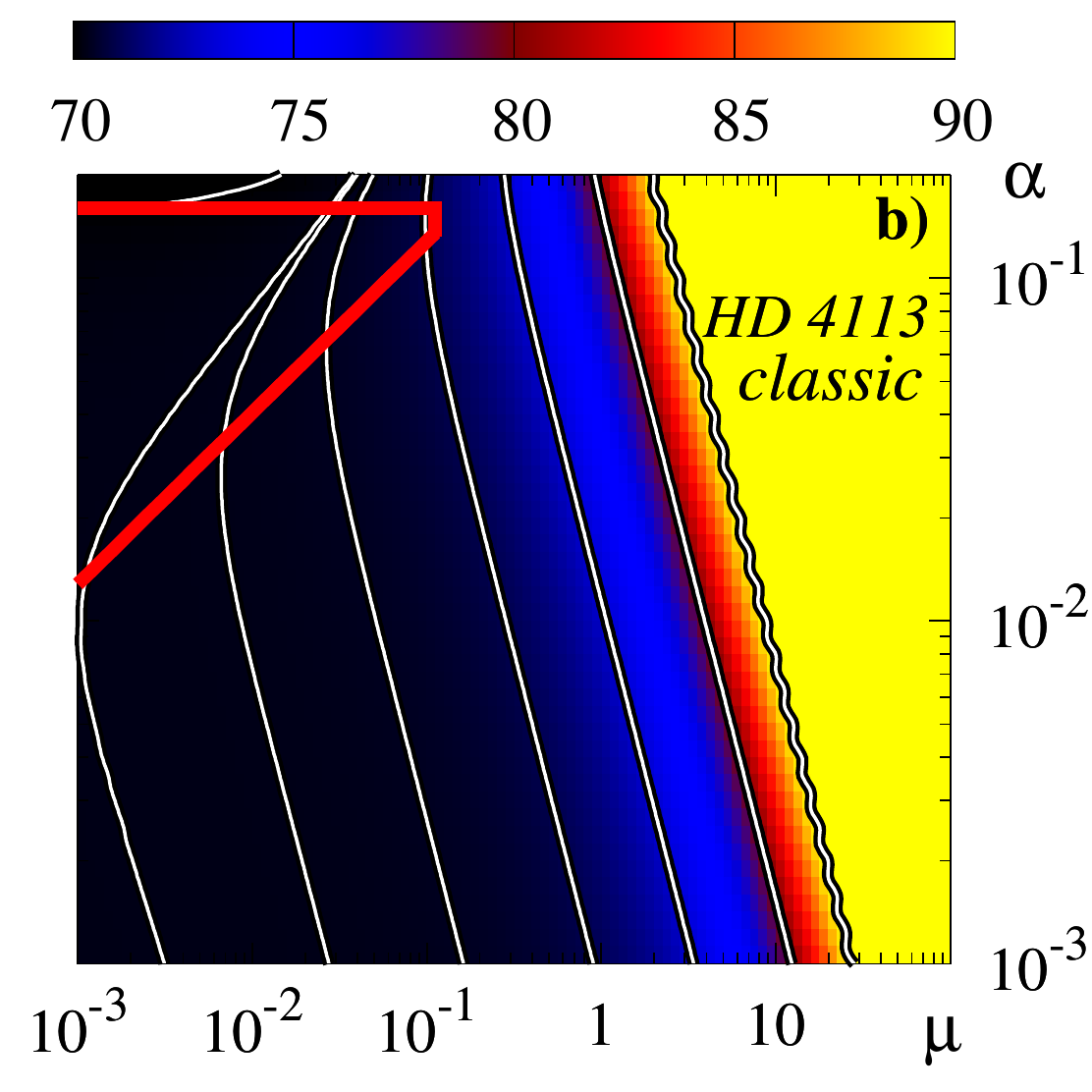}
          \includegraphics [width=4cm]{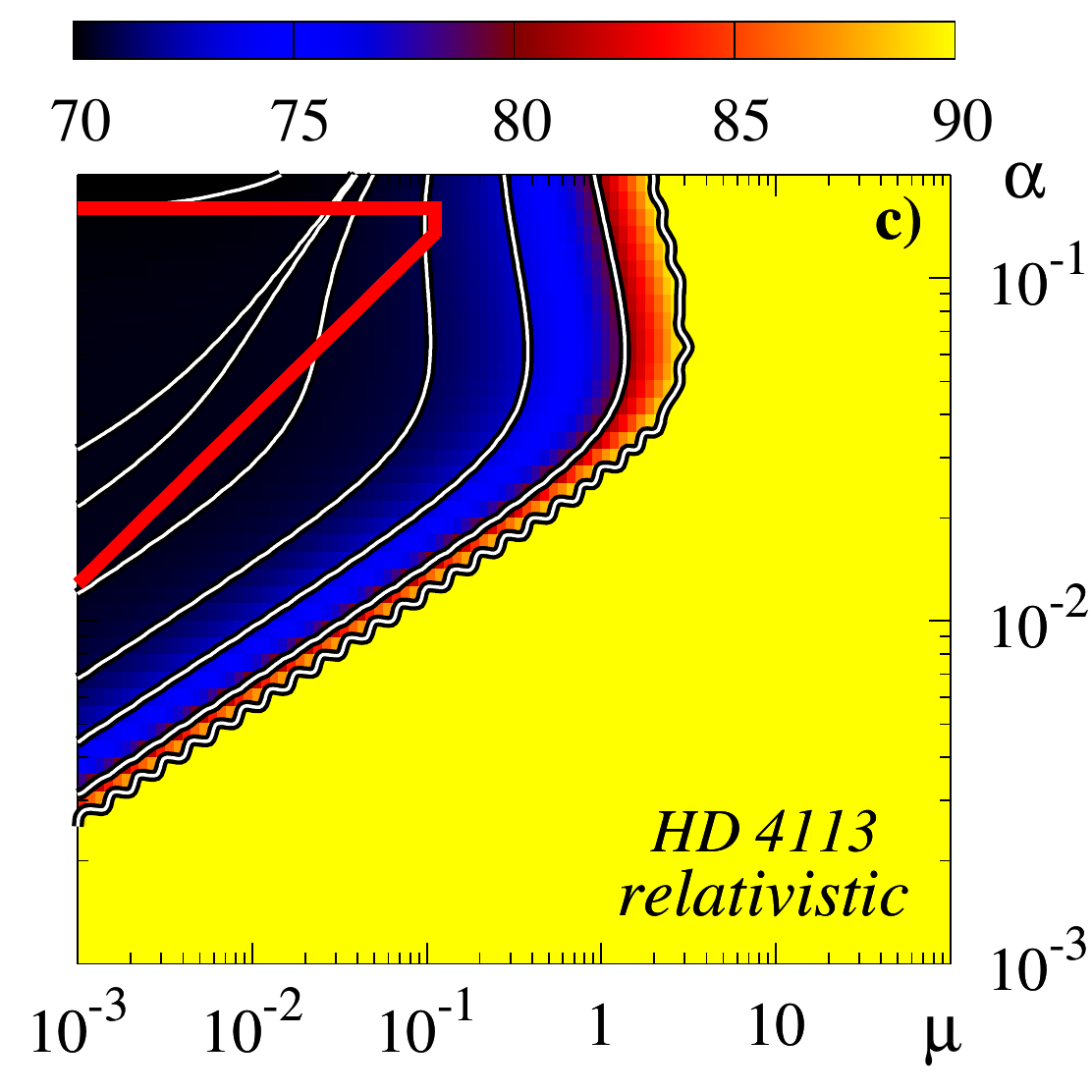}}
    \hbox{\includegraphics [width=4cm]{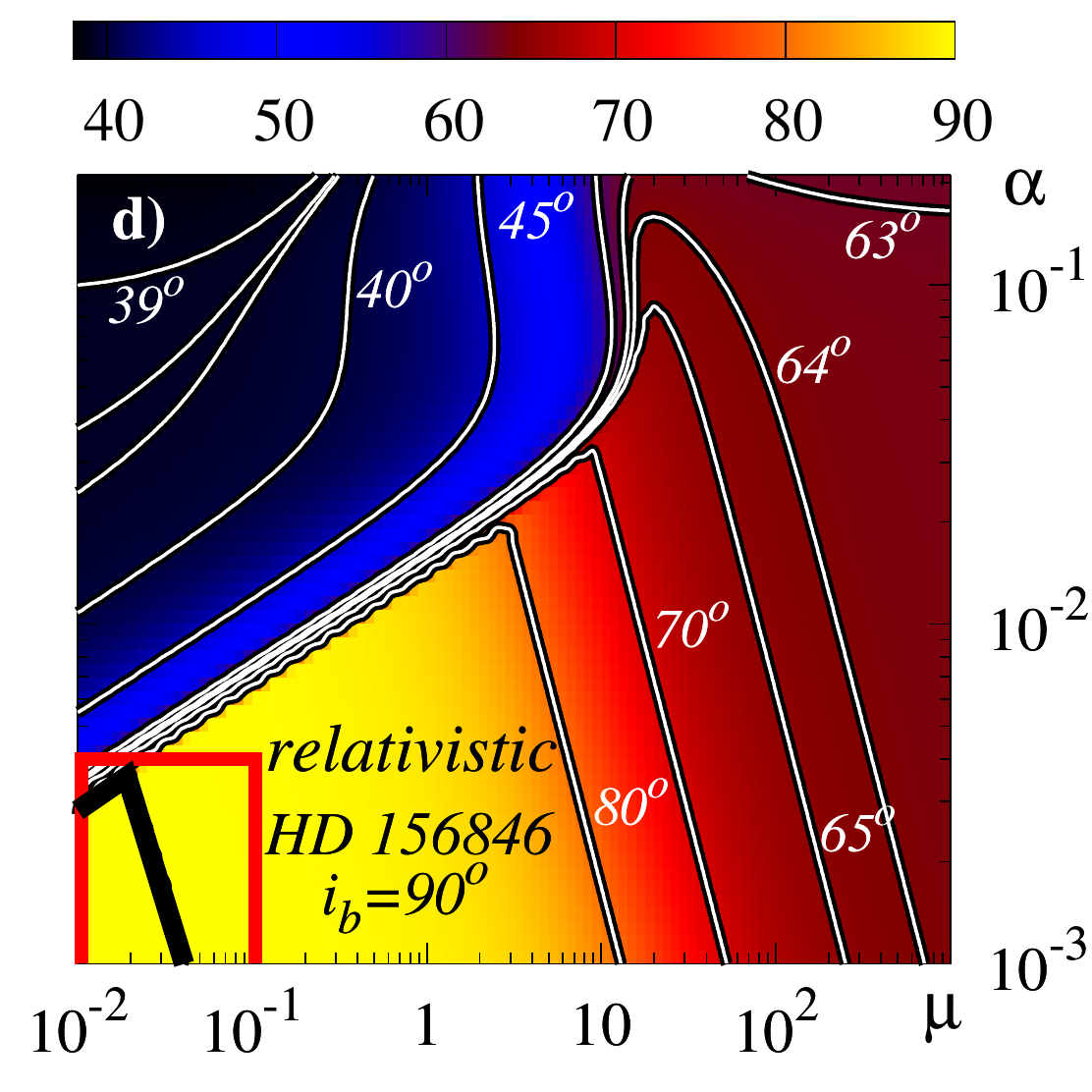}
          \includegraphics [width=4cm]{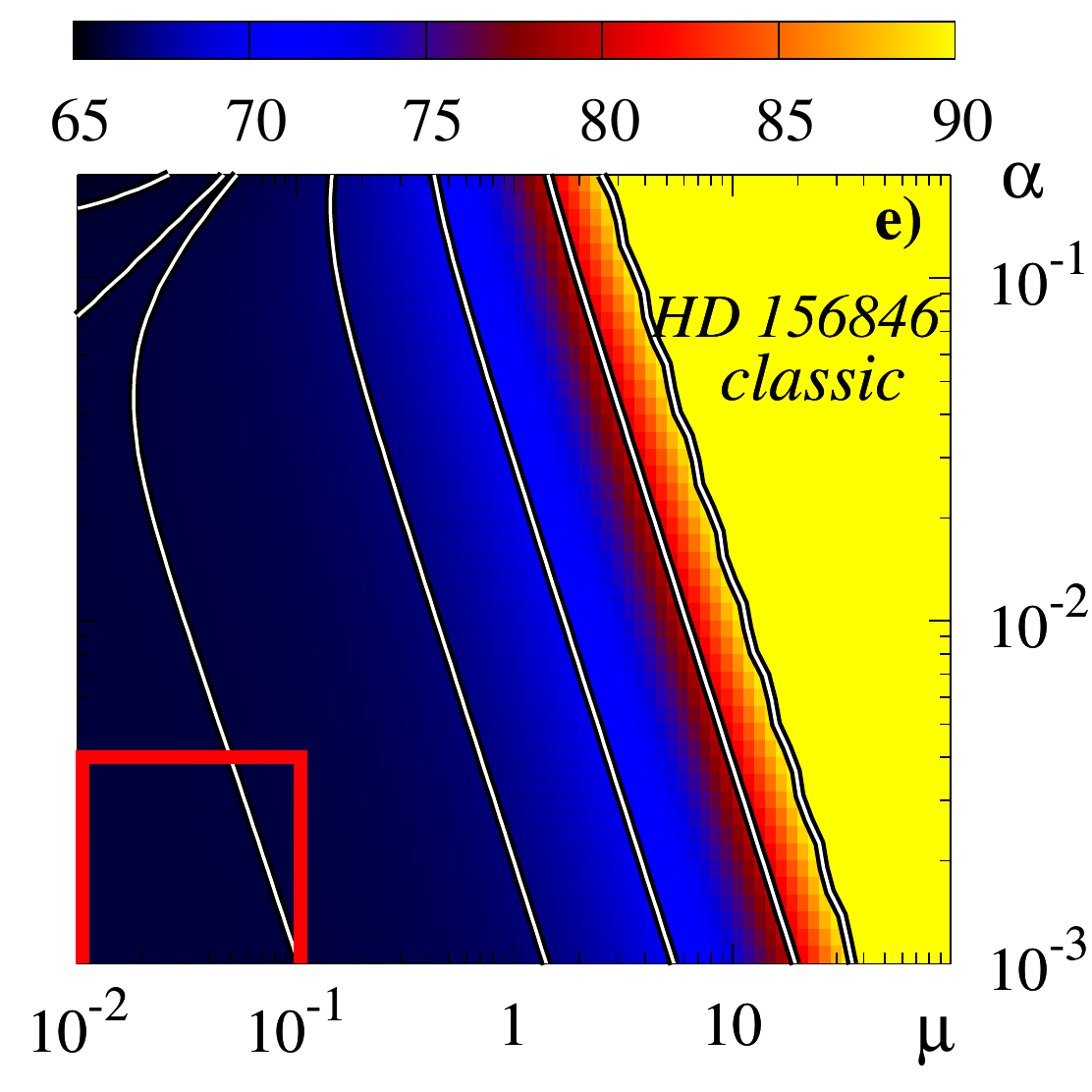}
          \includegraphics [width=4cm]{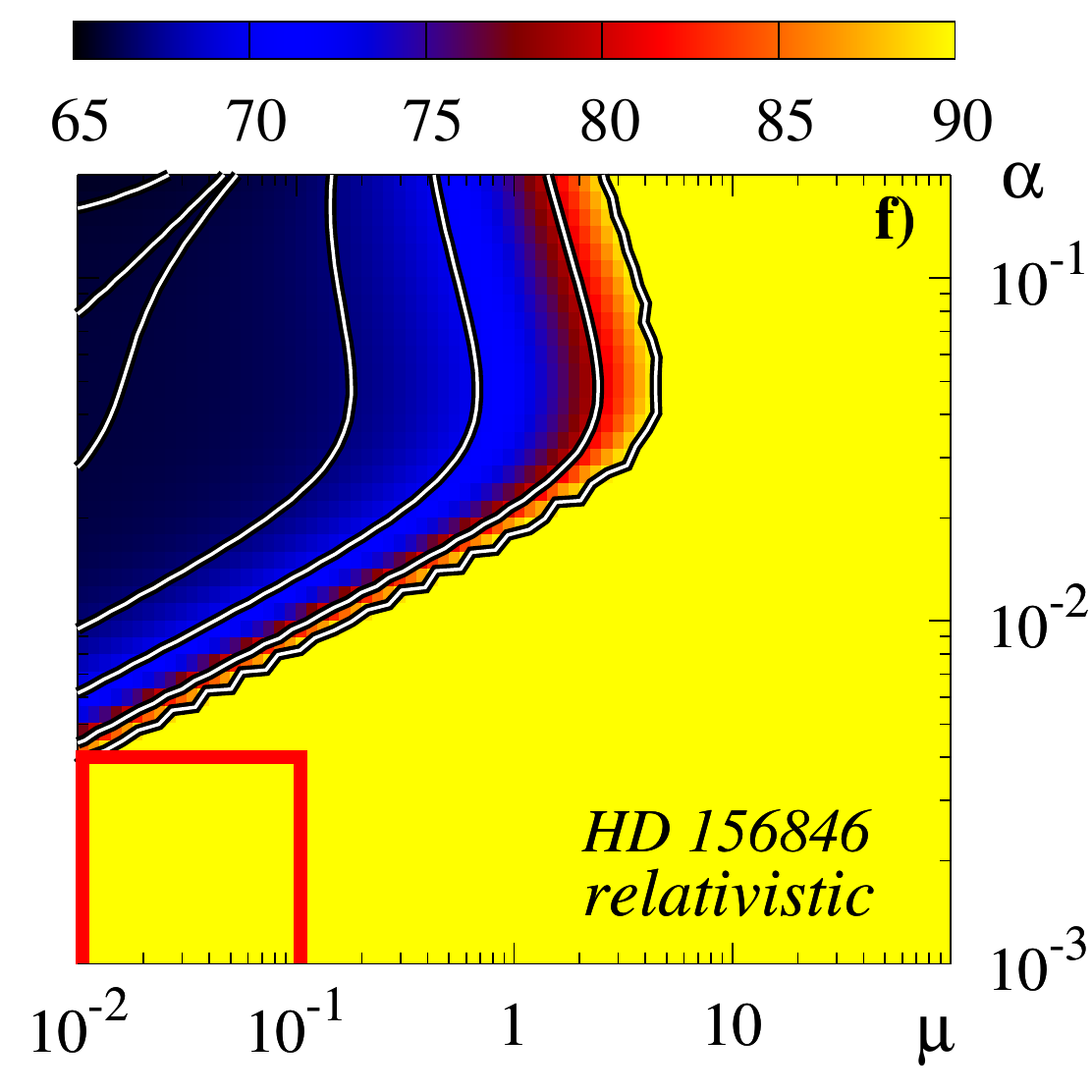}}	 
	 }
	 }
\caption{
The critical inclination $i_\idm{crit}$ ({\em the left-hand column}),  $\min
i_0(\max e)$ for the NG-model ({\em the middle column}), and  $\min i_0(\max e)$
for the GR-model ({\em the right-hand column}). The top row is for the HD~4133
system ($\max e = 0.9$), the bottom row is for the HD~156846 system ($\max e = 0.85$). 
}
\label{systems}
\end{figure}

%-------------------------------------------------------------------------------
%
We apply the results to test a hypothesis that highly eccentric orbits of
recently detected Jovian planets in HD~4113 and HD 156846 planetary systems
(\cite{Tamuz2008}) can be explained by the LKR resonance (in the sense
considered here, i.e., of the direct orbits) forced by more distant  and unseen
(likely massive) objects.  Indeed, the radial velocity (RV) of HD~4113 exhibits
annual trend  RV$_t \sim 28~\mbox{ms}^{-1}$ that implies the minimal mass of
putative distant companion  $\sim 10{\mbox{m}}_{\idm{J}}$ and $a_\idm{c} \sim
10\mbox{au}$. A simulation of $i_\idm{crit}$ is illustrated in
Fig.~\ref{systems}a. Here, we  assume that the orbit of HD~4113b is edge-on. The
red lines in the $(\mu,\alpha$)-plane mark raw limits of the orbital parameters
of a putative distant object. It  must be also massive enough to induce the
observed RV drift, hence the skew line is for the RV$_t$ estimated under an
assumption that its orbit is circular. Moreover, to force $\max e_\idm{b} \sim
0.9$ (to generate large enough ``loop'' in  the $\Sigma$-plane, see
Fig.~\ref{rel_rep}), appropriately large  $i_0 < \pi/2$, ($i_0 > i_\idm{crit}$),
is required. Figs.~\ref{systems}b,c  illustrate $\min i_0(\max e)$,
i.e., the minimal inclination $i_0$ for  which $e_1$ may reach given $\max e$
(here, $\max e = 0.9$).  Figure~\ref{systems}b is for the NG model and
Fig.~\ref{systems}c is for the GR model, respectively. In both cases, $\min
i_0(0.9) \sim 70^{\circ}$ and the putative body may be responsible
for the detected large eccentricity of HD~4113b.

The Jovian planet HD~156846b belongs to a wide binary with $a>250\mbox{au}$ and
$m_\idm{B} \sim 0.56{\mbox{m}}_{\odot}$. We found that the putative system is
located in such a $(\mu,\alpha)$-region,  in which the LKR can be suppressed at
all because $i_\idm{crit} \sim \pi/2$ (Fig.~\ref{systems}d). Moreover, $\min
i_0(0.85) \sim 66^{\circ}$ for the NG model (Fig.~\ref{systems}e), while $\min
i_0(0.85) > \pi/2$ for the GR model (Fig.~\ref{systems}f), also the structure  of the
$(\mu,\alpha)$-plane is qualitatively different in these two cases. Hence, in
the GR model, the  eccentricity cannot be explained by the LKR (in the sense
considered here). 

%
%-------------------------------------------------------------------------------
\section{Conclusions}
%-------------------------------------------------------------------------------
%
Recently (\cite{Migaszewski2009}), we found that apparently subtle GR correction
to the Newtonian model of coplanar planetary system may lead to significant,
qualitative changes of the secular dynamics. In the present work, we try to
extend such a quasi-global study to non-coplanar model, applying the averaging
and the concept of representative plane of initial conditions.  The results
indicate that the 3D dynamics are also very different in the both 3D models.  We
continue the work on this problem.
%
%-------------------------------------------------------------------------------
\section*{Acknowledgments}
%-------------------------------------------------------------------------------
%
This work is supported by the Polish Ministry of Science and Education,
Grant No. 1P03D-021-29. C.M. is also supported by Nicolaus Copernicus
University Grant No.~408A.

\bibliographystyle{astron}
\bibliography{ms}

\begin{thebibliography}{}

\bibitem[{{F{\'e}joz}}{~2002}]{Fejoz2002}
{F{\'e}joz}, J., 2002,
\newblock {\em Celestial Mechanics and Dynamical Astronomy,} {\bf 84}, 159

\bibitem[{{Ford} et~al.}{~2000}]{Ford2000}
{Ford}, E.~B., {Kozinsky}, B., and {Rasio}, F.~A., 2000,
\newblock {\em ApJ,} {\bf 535}, 385

\bibitem[{{Innanen} et~al.}{~1997}]{Innanen1997}
{Innanen}, K.~A., {Zheng}, J.~Q., {Mikkola}, S., and {Valtonen}, M.~J., 1997,
\newblock {\em AJ,} {\bf 113}, 1915

\bibitem[{{Kozai}}{~1962}]{Kozai1962}
{Kozai}, Y., 1962,
\newblock {\em AJ,} {\bf 67}, 579

\bibitem[{{Krasinsky}}{~1972}]{Krasinsky1972}
{Krasinsky}, G.~A., 1972,
\newblock {\em Celestial Mechanics,} {\bf 6}, 60

\bibitem[{{Krasinsky}}{~1974}]{Krasinsky1974}
{Krasinsky}, G.~A., 1974, Vol.~62 of {\em IAU Symposium,}, pp 95--116

\bibitem[{{Laskar} and {Robutel}}{~1995}]{Laskar1995}
{Laskar}, J. and {Robutel}, P., 1995,
\newblock {\em Celestial Mechanics and Dynamical Astronomy,} {\bf 62}, 193

\bibitem[{{Libert} and {Henrard}}{~2007}]{Libert2007}
{Libert}, A.-S. and {Henrard}, J., 2007,
\newblock {\em Icarus,} {\bf 191}, 469

\bibitem[{{Lidov} and {Ziglin}}{~1974}]{Lidov1974}
{Lidov}, M.~L. and {Ziglin}, S.~L., 1974,
\newblock {\em Celestial Mechanics,} {\bf 9}, 151

\bibitem[{{Michtchenko} et~al.}{~2006}]{Michtchenko2006}
{Michtchenko}, T.~A., {Ferraz-Mello}, S., and {Beaug{\'e}}, C., 2006,
\newblock {\em Icarus,} {\bf 181}, 555

\bibitem[{{Michtchenko} and {Malhotra}}{~2004}]{Michtchenko2004}
{Michtchenko}, T.~A. and {Malhotra}, R., 2004,
\newblock {\em Icarus,} {\bf 168}, 237

\bibitem[{{Migaszewski} and {Go{\'z}dziewski}}{~2008a}]{Migaszewski2008a}
{Migaszewski}, C. and {Go{\'z}dziewski}, K., 2008a,
\newblock {\em MNRAS,} {\bf 388}, 789

\bibitem[{{Migaszewski} and {Go{\'z}dziewski}}{~2008b}]{Migaszewski2008c}
{Migaszewski}, C. and {Go{\'z}dziewski}, K., 2008b,
\newblock {\em MNRAS, arXiv:0812.2949}

\bibitem[{{Migaszewski} and {Go{\'z}dziewski}}{~2009}]{Migaszewski2009}
{Migaszewski}, C. and {Go{\'z}dziewski}, K., 2009,
\newblock {\em MNRAS,} {\bf 392}, 2

\bibitem[{{Richardson} and {Kelly}}{~1988}]{Richardson1988}
{Richardson}, D.~L. and {Kelly}, T.~J., 1988,
\newblock {\em Celestial Mechanics,} {\bf 43}, 193

\bibitem[{{Takeda} and {Rasio}}{~2005}]{Takeda2005}
{Takeda}, G. and {Rasio}, F.~A., 2005,
\newblock {\em ApJ,} {\bf 627}, 1001

\bibitem[{{Tamuz} et~al.}{~2008}]{Tamuz2008}
{Tamuz} et~al., 2008,
\newblock {\em A\&A,} {\bf 480}, L33

\bibitem[{Verrier and Evans}{~2008}]{Verrier2008}
Verrier, P.~E. and Evans, N.~W., 2008,
\newblock {\em MNRAS, arXiv:0812.4528}

\end{thebibliography}
\label{lastpage}
\end{document}